\documentclass[12pt]{amsart}

\usepackage{times,amssymb,amsmath,float}
\usepackage{euscript} \usepackage[dvips]{epsfig}

  \newcommand{\probi}[1]{\mathbb{P}_i \left ( #1 \right
) } 

\newcommand{\ent}[1]{h\! \left (\! #1 \! \right)}
\newtheorem{proposition}{Proposition}
\newtheorem{theorem}[proposition]{Theorem}

\newtheorem{lemma}[proposition]{Lemma}

\newcommand{\remove}[1]{} \newcommand\nc\newcommand
\newcommand\nd{\noindent}%\newcommand\bysame{\rule[1mm]{1cm}{.025cm}}

 \nc\dgv{\delta_{\text{\rm GV}}}
\nc\dlp{\delta_{\text{\rm LP}}} \nc\rcrit{R_{\text{crit}}}
\renewcommand\epsilon{\varepsilon}

\newcommand{\beeq}{\begin{eqnarray*}}
\newcommand{\eneq}{\end{eqnarray*}}

\begin{document}
\title[] {On the asymptotic accuracy of the union bound}
\thanks{} 

\author[]{Alexander Barg$^{\ast}$} 
\thanks{$^{\ast}$ 
Dept. of ECE, University of Maryland,  College Park, MD 20742. E-mail
{abarg@eng.umd.edu}.  Research
supported in part by NSF Grant CCR-0310961.}  

\begin{abstract} A new 
lower bound on the error probability of maximum likelihood decoding
of a binary code on a binary symmetric channel (BSC) was
proved in Barg and McGregor (2004, cs.IT/0407011). It was observed in that
paper that this bound leads to a new region of code rates in
which the random coding exponent is asymptotically tight, giving
a new region in which the reliability of the BSC is known exactly.
The present paper explains a relation of these results to the union
bound on the error probability.
\end{abstract}

\maketitle
\thispagestyle{empty}

\section{Introduction}
This is a companion paper to \cite{bar04b}. Suppose that a code $C$
is used on a BSC$(p)$ and decoded according to the maximum likelihood
procedure. The error probability of decoding $P_e(C,p)$ can be estimated from
above using the distance distribution of $C$ together with the union bound.
As a general rule of thumb, this bound gives a good estimate of the 
error probability for low channel noise and is loose for high noise.
Quantifying this heuristic is a difficult problem 
related not just to the distance distribution but also
to structural properties of the code. 
Rigorous results are attainable
only in the asymptotic setting when the code length $n$ tends to infinity
(therefore in effect we will study families of codes rather than
individual codes without always saying so). 
The inaccuracy of the union bound is related to the fact that 
intersections of half-spaces related to codewords other than the
transmitted one, are counted more than once.
It turns out that under certain conditions adding the measure of
these intersections does not change the exponential asymptotics of the actual
value of the error probability. The first result of this type
was obtained by Gallager \cite{gal73} who proved that
for the ensemble of random codes and for rate $R<\rcrit,$ where $\rcrit$
is the so-called critical rate of the channel (see below), 
the union bound gives the correct exponent of the average
error probability for this ensemble (this quantity is different from the error
probability of a typical random code, and both are different 
the error probability of decoding for a typical linear code,
see \cite{bar02b}). The proof in \cite{gal73} is based on the fact that
the error probability of decoding into a list of size
two decreases exponentially faster than the estimate of $P_e(C,p)$
given by the union bound. A similar result can be proved for
the ensemble of random linear codes using the ensemble-average coset
weight distribution.

Subsequent results of this type are substantially more involved.
They are related to universal bounds on the distance distribution
of codes \cite{lit99,ash99a} and rely upon various methods of proving
lower bounds on $P_e$ given the distance distribution. One such method,
due to \cite{kou68}, was used in \cite{lit99,ash00c} to prove new
estimates of the reliability function of the BSC \cite{lit99} and
the power-constrained AWGN channel \cite{ash00c}.
Other methods known are due to \cite{bur84,bur00} and \cite{coh04}.
The main question addressed by this analysis is the value of the
code rate $R_\ast$ such that for rates $R\le R_\ast$ the union
bound can be claimed to be exponentially tight. 

The paper is organized as follows. In Sect.\,\ref{sect:problem} 
we discuss the problem statement. 
Sect.\,\ref{sect:bounds} is devoted to general lower estimates of the error
probability $P_e(C,p)$ given the distance distribution of the code $C$.
In Sect.\,\ref{sect:geometry} 
we study the relation between the random coding exponent
(the exponent of the error probability for a typical linear code)
and the union bounds on this probability. Our context is that of
geometry of decoding of random linear codes. We explain how
different bounds on codes are related to the union bound on the error
probability. Then in Sect.\,\ref{sect:results} we put everything 
together and show that a part of the random coding exponent just
below the critical rate of the channel gives the actual value of the
channel reliability. Some concluding remarks are presented in the 
final Section \ref{sect:conclude} 

\section{Statement of the problem}\label{sect:problem}

We consider transmission with binary codes of length $n$ over a BSC
with crossover probability $p$.  Let $X=\{0,1\}^n$
be the $n$-dimensional Hamming space.  Let $C(n,M=2^{Rn})\subset X$ be
a code of rate $R$ and let $x_i\in C$ be the transmitted vector. Under
this condition the probability that a vector $y$ is received equals
$P(y|x_i)=p^{|y+x_i|}(1-p)^{n-|y+x_i|},$ where $|\cdot|$ is the
Hamming weight.  
%For a given set $S\subset X$ let $\probi{S}=\sum_{y\in S}P(y|x_i).$

Let $D(x)$ be the decision region of max-likelihood decoding for a
codevector $x$.
%For a vector $x\in C$ define its Voronoi region $D(x)$ as follows:
%\[
%D(x)=\{y\in X: \forall_{x'\in C\backslash x}\dist(x,y)<\dist(x',y)\}.
%\]
Given that $x_i$ is transmitted, the error probability of maximum
likelihood decoding equals  $P_e(x_i)=\probi {X\backslash D(x_i)}.$
The (average) error probability of decoding for the code $C$ equals
\[
P_e(C,p)=\frac 1M \sum_{i=1}^M P_e(x_i).
\]
Computing this probability directly is prohibitively difficult in most
nontrivial examples, therefore, there has been much interest in
bounding it from both sides. As in \cite{bar04b}, we
focus on {\em lower bounds} on $P_e(C,p).$ 
%Other recent papers devoted to this problem include
%\cite{ash00c,bur00,coh04,ker00,kua00b,lit99,seg98}.
For a given code sequence, define its error
exponent as
  $$
    E(p)=\lim_{n\to\infty}\frac 1n \log \frac 1{P_e(C,p)}.
  $$
We will also apply the results of the paper to the largest attainable
exponent of the error probability of decoding defined as
  $$E(R,p)=\limsup_{n\to\infty} \frac{1}{n} \log \max_{C\subseteq
X, R(C)=R} \frac 1{P_e (C,p) }.
  $$
This quantity is also called the {\em reliability function} of the BSC.

Let us fix an arbitrary ordering of the codewords.
Define the {\em local} distance distribution of the code $C$ with respect
to the codeword $x_i$. This is a set of $n+1$ numbers $B_0^i,\dots,B_w^i,
\dots,B_n^i$, where $B_w^i$ is the number of neighbors of $x_i$ in the 
code at distance $w$.
Below we will mostly concentrate on lower bounds on the probability
$P_e(x_i)$ given the {local} distance distribution.
We will consider codes of exponentially growing size
for which the error probability $P_e(C,p)$ declines exponentially fast. 
In this situation, given the {\em average} distance distribution
of the code $C$, we can isolate a subcode of the same exponential
order in which the local distance distribution for every codeword
is asymptotically the same as the average distribution. Therefore,
the bound $P_e(x_i)$ can be used to obtain a bound on $P_e(C,p)$
with the same exponent. This argument is presented in detail in 
\cite{ash00c, bur00}, so we will rely on it here without further discussion.

\subsection*{Notation} Let $C=\{x_1,\dots,x_M\}$
 be a code.
For a subset $Y\subset X$ let
   $$
    \probi Y=\sum_{y\in Y}P(y|x_i).
   $$
Let $\pi(w)$ be the error probability for two codewords at distance
$w$, i.e., the probability of transmitting $x_i$ and decoding $x_j$,
where $d(x_i,x_j)=w$ and $d(\cdot,\cdot)$ denotes the Hamming distance.
By the union bound,
   \begin{equation}\label{eq:union}
     P_e(x_i)\le \sum_{w=1}^n B_w^i \pi(w).
   \end{equation}
Letting $\pi(\omega n)=2^{n A(\omega)+o(n)}$, we have 
$A(\omega)=\omega\log2\sqrt{p(1-p)}.$ Then 
   \begin{equation}\label{eq:union-exp}
    \frac 1n\log\frac 1{P_e(x_i)}\gtrsim -A(\omega)-\mu(\omega),
   \end{equation}
where $\mu(\omega)=\frac1n\log B_w^i.$ 

By $h(x)$ we denote the binary entropy function. We also use the divergence
$D(x\|y)=h(x)+x\log y+(1-x)\log(1-y)$ (the logarithms are binary).

\subsection*{Bounds on codes}
Define
   $$
     \delta(R)=\limsup_{n\to\infty} \max_{C: |C|=2^{nR}} \frac {d(C)}{n}
   $$
There exist code sequences
(for instance, typical codes from the ensemble of random linear codes)
whose relative distance approaches the quantity $\dgv(R)=h^{-1}(1-R)$ which 
is called the Gilbert-Varshamov (GV) distance. Thus,
  $$
    \delta(R)\ge \dgv(R).
  $$
On the other hand, by the Elias bound,
  $$
    \delta(R)\le \delta_E(R):=2\dgv(R)(1-\dgv(R))
  $$
where the quantity $\delta_E(R)$ is sometimes called the Elias distance.
A better upper estimate of $\delta(R)$ is provided by the JPL bound
\cite{mce77a}:
\[
\delta(R)\le \bar\delta :=\min_{\substack {0\le\alpha\le \frac12}}
G(\alpha,\tau)
\]
where $G(\alpha,\tau)=2\frac{\alpha(1-\alpha)-\tau(1-\tau)}
{1+2\sqrt{\tau(1-\tau)}}$, and where $\tau$ satisfies
$h(\tau)=1-R-h(\alpha).$
For $0\le R\le 0.305$ this bound takes a simpler form:
  $
    \bar\delta= \phi(h^{-1}(R)),
  $
where $\phi(x)=\frac12 -\sqrt{x(1-x)}.$
Denote by $\bar R(\delta)$ the inverse function of $\bar\delta(R)$
which is well defined because $\bar\delta$ is a monotone decreasing
function of $R$.

\section{Lower bounds on $P_e(C,p)$}\label{sect:bounds}

In this section we review the known lower estimates of the probability
$P_e(x_i)$ given the local distance distribution of the code.
Let $C(i)=\{x\in C\,:\,d(x,x_i)=w\}$
for some fixed value of $w.$ 
Given two different vectors $x_i,x_j\in C$, let 
        $$X_{ij}\subset\tilde X_{ij}:=\{y\in X\,:\, d(x_j,y)\le d(x_i,y)\}
     $$ be an arbitrary subset.
\subsection{Kounias' bound \cite{kou68}} This (obvious) bound states that
\begin{equation*}\label{eq:kounias} P_e(x_i) \ge \sum_{x_j\in
C(i)} \Big\{\probi {X_{ij}}-\sum_{\begin{substack}
{x_k\in C(i)\backslash\{x_j\}\\k<j}\end{substack}} \probi
{X_{ij}\cap X_{ik}}\Big\}.
\end{equation*} 

In principle, here and hereafter $C(i)$ can be an arbitrary subcode of $C$
that does not contain $x_i$.

\subsection{Burnashev's method \cite{bur84,bur01b,bar04b}}
This method was originally suggested for the AWGN channel and
was adapted to the BSC in \cite{bar04b}. 
The error probability of decoding is estimated by carefully 
taking account of
the probability of the subsets $X_{ij}\cap X_{ik}, k\ne j$ for 
$x_j,x_k\in C(i)$ and for some suitable definition of the subsets
$X_{ij}.$ 
Let $x_i,x_j,x_k\in C(i), d(x_i,x_j)=d(x_i,x_k)=\omega n, 
d(x_j,x_k)=\lambda n.$ Let 
  \begin{equation}\label{eq:Xij}
    X_{ij}=\{y \in X : d(x_i,y)=d(x_j,y)= \frac{\omega n}{2}+pn(1-\omega)\}.
  \end{equation}
Denote by $B(\omega,\lambda)$ the negative exponent of the probability
$\probi {X_{ik}|X_{ij}},$ 
\begin{multline}\label{eq:Bwl}
B(\omega,\lambda) = -\omega-(1-\omega)h(p)+ \\ \max_{\eta\in
[\frac{\lambda p}{2},\min(\frac{\lambda}{4},p(1-\omega))]} \left (
\lambda \ent{\frac{2\eta}{\lambda}}+(\omega-\lambda/2)
\ent{\frac{\omega-2\eta}{2\omega-\lambda}}+
(1-\omega-\lambda/2)\ent{\frac{p(1-\omega)-\eta}{1-\omega-\lambda/2}}
\right ).
\end{multline}
The main result of \cite{bar04b} is given by
\begin{theorem}\label{thm:new}\cite{bar04b} 
 Let $(C_i)_{i\ge 1}$ be a sequence of codes
with rate $R$, relative distance $\delta$ 
and distance distribution satisfying $B_{\omega n}\ge
2^{n\beta(\omega)-o(n)},$ where $\beta(\omega)>0$ for all 
$\delta\le\omega\le 1.$ 
The error probability of max-likelihood
decoding of these codes satisfies $P_e(C,p)\ge 2^{-En+o(n)},$ where
\begin{equation}\label{eq:new}
E= \min_{\delta\le\omega\le 1}\; \max_{0\le \lambda\le \omega}\;
\big[\max (-\beta (\omega) -A(\omega) ,  B(\omega,\lambda)
-A(\lambda))\big].
\end{equation}
%where $A$ and $B$ are defined as in Equations {\rm(\ref{eq:Aw})} and
%{\rm (\ref{eq:Bwl})} respectively.
\end{theorem}

As it turns out, for sufficiently low code rates $R$, the
first term under the maximum in (\ref{eq:new}) dominates the estimate.
This shows that for code rates $R\le R_\ast$ the union bound 
is exponentially tight, where $R_\ast$ is some value of the rate
than depends on the distance distribution of the
code and on the noise level in the channel.
We will study the values of $R_\ast$ in Sect.\,\ref{sect:results}
for the problem  of bounding the channel reliability function.

\subsection{The method of Cohen and Merhav: de Caen's inequality
and its generalizations}
D. de Caen \cite{cae97} suggested a new lower bound
on the probability of a finite union of events. While an elementary
result (essentially, Cauchy-Schwarz), this bound is sometimes the best
among the inequalities of this type. De Caen's inequality was used to
compute lower bounds on the error probability via the distance
distribution in \cite{seg98,ker00}. Cohen and Merhav \cite{coh04}
generalized de Caen's inequality by introducing a weighting function
that depends on the weight of the error vector
 and derived a lower bound on $P_e(C,p)$ by optimizing on this function.
Their result can be stated as follows.
\begin{theorem}\label{thm:cm}\cite{coh04} 
Let $x_j,x_k\in C(i)$ be arbitrary vectors, $j\ne k.$ 
Then
\begin{equation} \label{eq:cm}
P_e(x_i)\ge 
\frac{B_w^i\Big[\sum\limits_{y\in \tilde X_{ij}}P(y|x_i)\eta(|y|)\Big]^2}
{\sum\limits_{y\in \tilde X_{ij}}P(y|x_i)\eta^2(|y|)+(B_w^i-1)
\sum\limits_{y\in \tilde X_{ij}\cap \tilde X_{ik}}P(y|x_i)\eta^2(|y|)},
\end{equation}
where
$\eta(\cdot)$ is an arbitrary  weight function.
\end{theorem}
Taking $C(i)$ to be the set of neighbors of $x_i$ at the minimum distance
$d$, 
paper \cite{coh04} obtains a bound on $P_e(x_i)$ formed of two pieces.
Similarly to Theorem \ref{thm:new},
Theorem \ref{thm:cm} implies that for low rates the exponent of 
$P_e(x_i)$ asymptotically coincides with the exponent of the union bound.
The condition on the code rate for the union bound on $P_e(x_i)$
to be (exponentially)
tight proved in \cite[Prop.\,5.3]{coh04} can be written as follows;
   \begin{equation}\label{eq:condition}
        B_d^i \probi{X_{ij}\cap X_{ik}}\lesssim \probi{X_{ij}},
   \end{equation}
where $x_j,x_k\in C(i)$ are arbitrary (different) codewords and
$\lesssim$ refers to an inequality for the exponents\footnote{Note that
(\ref{eq:condition}) relies on $X_{ij}$ instead of $\tilde X_{ij}$. The reason
for this is explained in the end of Sect.\,\ref{sect:results} below.}.

\section{Decoding geometry of random linear codes 
and the union bound}\label{sect:geometry}

\subsection{Decoding of random linear codes}
Consider the ensemble of linear codes defined by $(n-k)\times n$
parity-check matrices with independent random components chosen
with equal probability from $\{0,1\}.$ Let $R=k/n.$
The ensemble-average weight distribution has the form $A_{\omega n}
\cong 2^{n(R+1-h(\omega))}, \omega=0,(1/n),\dots,(n-1)/n,1.$ 
The minimum relative distance $\delta$ of a typical code from the ensemble 
approaches the Gilbert-Varshamov bound $\dgv(R)=h^{-1}(1-R).$
Computing the error probability $P_e(C)$ for such a code, we obtain
an upper bound on the BSC reliability of the form $E(R,p)\ge E_0(R,p),$
where $E_0(R,p)$ is the ``random coding exponent,''
\begin{equation}\label{eq:Ge}
E_0(R,p)=\left\{\begin{array}{l@{\quad}l@{\qquad}c} -\dgv(R) \log_2
2\sqrt{p(1-p)} &0\le R\le R_x, &{\rm (a)}\\[2mm]
D(\rho_0\|p)+\rcrit-R%1-R-\log_2(1+2\sqrt{p(1-p)})
  &R_x \le R\le
\rcrit ,  &{\rm (b)}\\[2mm] D(\dgv(R)\|p) &\rcrit\le R\le1-h (p),
&{\rm (c)}
\end{array}\right.
\end{equation}
where
\begin{eqnarray}
R_{x}&=&1-h_2 (\omega_0)
%\Big(\frac{2\sqrt{p(1-p)}}{1+2\sqrt{p(1-p)}}\Big)
\label{eq:r-exp}\\
\rcrit&=&1-h_2(\rho_0)%\Big(\frac {\sqrt p}{\sqrt p+\sqrt{1-p}}\Big).
\label{eq:r-crit}
\end{eqnarray}
\begin{equation}\label{eq:rho0}
\rho_0=\frac {\sqrt p}{\sqrt p+\sqrt{1-p}},
\qquad\omega_0:=2\rho_0(1-\rho_0)=
\frac{2\sqrt{p(1-p)}}{1+2\sqrt{p(1-p)}}.
\end{equation}
This is a classical result of coding theory due to P. Elias and R. Gallager.
Concise, self-contained proofs that are suitable for our context appear
in \cite{bar02b,bar02h}. 

A part of this result that is used below is related to the typical weight 
$\omega_{\text{typ}} n$
of the incorrectly decoded codeword in the case of decoding error\footnote{The
expression for $P_e(C)$ is a finite sum of binomial-type
probabilities. 
Asymptotically for large $n$ it is dominated by weights of incorrectly
decoded codewords in a small segment around some value, which is 
called a {\em typical} weight of incorrect codewords.}. For the
cases (a)-(c) of (\ref{eq:Ge}) the
values of $\omega_{\text{typ}}$ are as follows \cite{bar02b}:
  \begin{alignat*}{2}
     &\text{(a)}  &&\omega_{\text{typ}}=\dgv(R)\\
     &\text{(b)}  \quad&&\omega_{\text{typ}}=\omega_0\\
     &\text{(c)}  &&\omega_{\text{typ}}=\delta_E(R).
  \end{alignat*}

\begin{figure}[t]
\epsfxsize=9cm \setlength{\unitlength}{1cm}
\begin{center}
\begin{picture}(6,6)
\put(-1.7,2){\epsffile{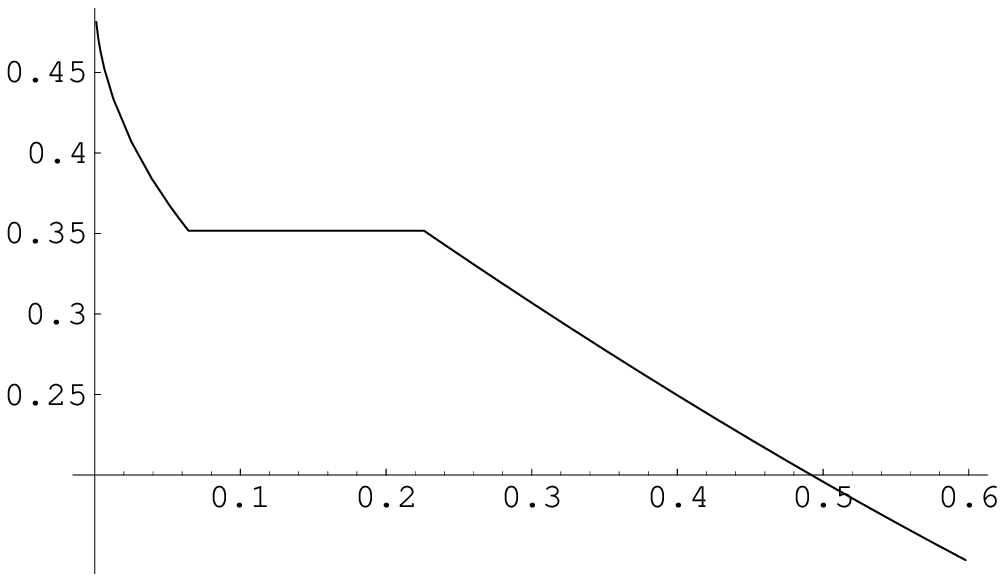}}
\put(-1.5,-3.2){\epsffile{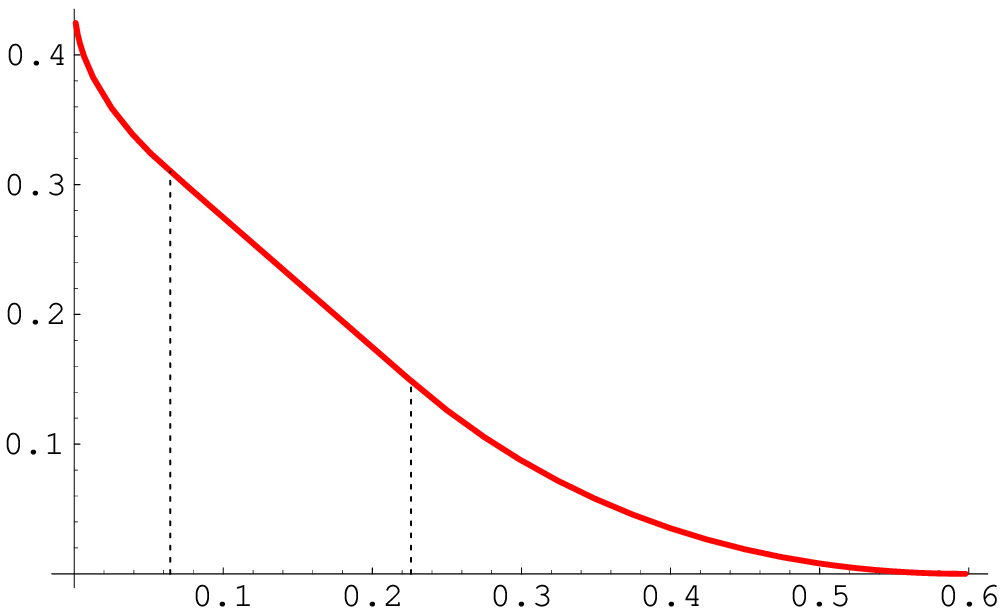}}
\put(-.4,6.3){{\footnotesize\mbox{$\dgv(R)$}}}
\put(.8,5.3){{\footnotesize\mbox{$\omega_0$}}}
\put(3.7,4.2){{\footnotesize\mbox{$\delta_E(R)$}}}
\put(-1.6,7.2){{\footnotesize\mbox{$\omega_{\text{typ}}$}}}
\put(-2.3,2.2){{\footnotesize\mbox{$E_0(R,p)$}}} 
\put(6.7,3.2){{\footnotesize\mbox{$R$}}}
\put(6.7,-2.5){{\footnotesize\mbox{$R$}}}
\put(0.1,-2.5){{\footnotesize\mbox{$R_x$}}}
\put(1.5,-2.5){{\footnotesize\mbox{$\rcrit$}}}
\end{picture}\end{center}
\vskip3.5cm
\caption{The typical weight of incorrect codewords and the random
coding exponent for a BSC with
$p=0.08$. }\label{fig:errbnds}
\end{figure} 

In Fig.\,\ref{fig:errbnds} the bound $E_0(R,p)$ is shown together with
the values $\omega_{\text{typ}}$ as a function of the code rate $R$.
As $R$ varies between $R_x$ and $\rcrit$, the value of $\omega_{\text{typ}}=
\omega_0$ changes its location with respect to the minimum
distance of the code, moving from $\dgv(R)$ to $\delta_E(R)$.
We note that $\omega_{\text{typ}}<\delta_E(R)$ as long as $R\le\rcrit.$ 

\subsection{Weight distributions and the union bound on $P_c(x_i)$} 
It is conjectured that $E_0(R,p)$
gives an exact value of $E(R,p)$ for all $R\in [0,1-h(p)].$ In an attempt
to prove this, various upper bounds on $E(R,p)$ were established.
The tightest known upper bounds are proved by showing that an appropriate
version of the union bound in effect is tight 
(entails no loss of accuracy of the estimate for large $n$).

The weight profile (the exponent of the weight distribution) 
of a typical random linear code of rate $R$
has the form $R+1-h(\omega), \omega\ge \dgv(R).$ As explained above,
only the weights in the region $\dgv(R)\le \omega\le \delta_E(R)$ are
relevant for the random coding exponent. Let us assume for a moment
that  

\smallskip\nd
(A) for any code $C$, a given codeword $x_i$ has at least
$2^{n(R+1-h(\omega))}$ codeword neighbors at relative distance 
$\omega=g(R)$ were $g$ is some monotone decreasing function;

\smallskip\nd (B) the union bound gives a tight value of
the error exponent in the estimates (\ref{eq:new}) and/or (\ref{eq:cm})
for some region of low rates, to be specified later.

\smallskip\nd By (B), we can write an asymptotic estimate of
$P_e(x_i)$ using (\ref{eq:union} in the reverse direction.
Substituting the distance distribution from (A)
we would be able to state an upper bound on $E(R,p)$ of the 
form 
\begin{equation}\label{eq:rub}
E(R,p)\le -(R-1+h(g(R)))-A(g(R)).
\end{equation}

For instance, if (A) were true for $\omega=\dgv(R)$ then 
we would obtain (\ref{eq:Ge}a) as an {\em upper} bound on $E(R,p)$
(this is a very strong assumption because it implies that the GV bound 
is tight). In this case $g(R)=\dgv(R)$.

 We will assume that $g(x)$ is
such that the function $-(R-1+h(g(R)))-A(g(R))$ is $\cup$-convex
(this will be the case in all our examples).
%In Fig.\,\ref{fig:union} we show a few ``union bound exponents''
%together with the straight-line bound (\ref{eq:Ge}b). 

Two important
remarks should be made with respect to this argument and 
Fig.\,\ref{fig:rel}.
We formulate the first one as
\begin{lemma}\label{lemma:tangent}
  The function on the right-hand side of 
(\ref{eq:rub}) is tangent to the straight line $D(\rho_0||p)+\rcrit-R$ at the 
point $R_1=g^{-1}(\omega_{\rm typ}).$
\end{lemma}
Thus if $g^{-1}(\omega_{\rm typ})< \rcrit$, 
the random coding bound $E_0(R,p)$ of (\ref{eq:Ge})
gives an exact answer for the channel reliability $E(R,p)$ 
at the point $R=R_1.$ Furthermore, together with the straight-line principle
of \cite{sha67} this implies that $E(R,p)=E_0(R,p)$ 
for all rates $R_1\le R\le \rcrit.$ A result of this type will be
proved in the next section.

Secondly, if $\omega_{\rm typ}=\delta_E(R)$ then it turns out that almost
every error vector from the sphere of typical errors leads to a decoding
error (see e.g., \cite{bar02h}). Therefore, for $R\ge \rcrit$
instead of (\ref{eq:rub}) we compute a ``union bound'' of a different type,
namely, the probability of an error vector of weight $\dgv(R)$ occurring in 
the channel. This argument is not related to the above assumptions and
gives (\ref{eq:Ge}c) as an unconditional upper bound on $E(R,p)$
(the sphere-packing bound).
\remove{\begin{figure}[t]
\epsfxsize=6cm \setlength{\unitlength}{1cm}
\begin{center}
\begin{picture}(6,6)
\put(0,0){\epsffile{ubs.eps}}
\end{picture}\end{center}
\vskip0cm
\caption{Union bounds for $E(R,p)$}\label{fig:union}
\end{figure} }

\section{Reliability function of the BSC}\label{sect:results}

\begin{figure}[tH]
\vspace*{1cm}
\epsfxsize=10cm \setlength{\unitlength}{1cm}
\begin{center}
\begin{picture}(6,6)
\put(-1,0){\epsffile{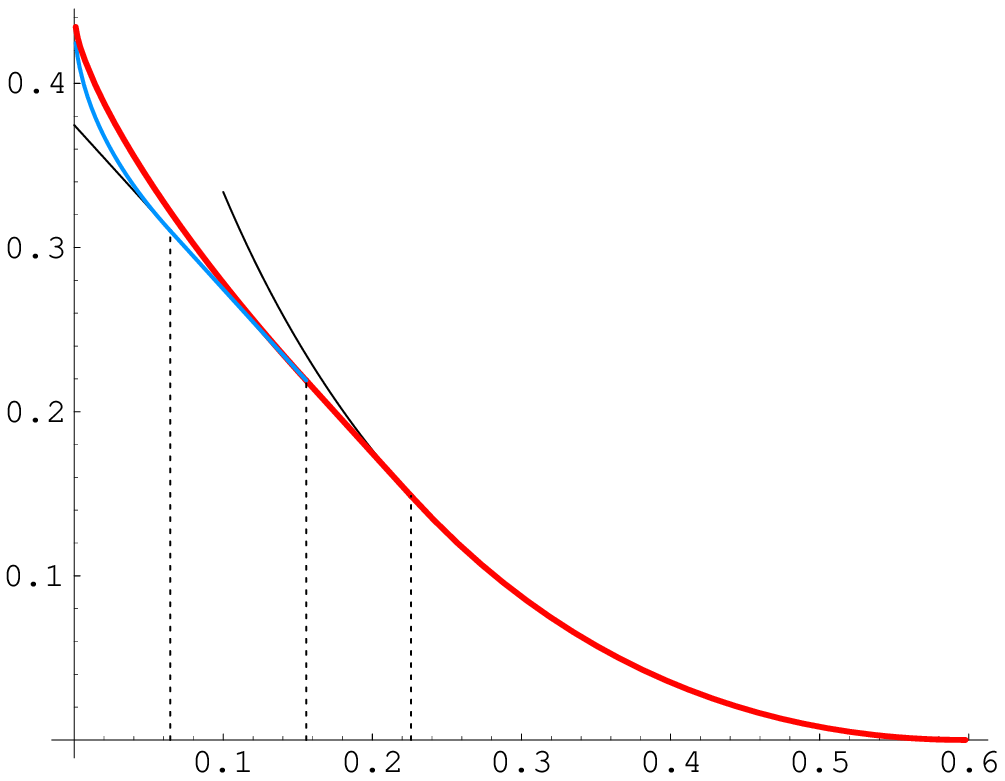}} 
\put(8.7,0.7){{\footnotesize\mbox{$R$}}}
\put(0.2,0.7){{\footnotesize\mbox{$R_x$}}}
\put(-1.5,7.5){{\footnotesize\mbox{$E(R,p)$}}}
\put(1.6,0.7){{\footnotesize\mbox{$R_1$}}}
\put(3.3,0.7){{\footnotesize\mbox{$\rcrit$}}}
\end{picture}\end{center}
\caption{Bounds on the error exponent for the BSC with
$p=0.08$. In the interval $R_1\le R\le \rcrit$ the random
coding bound $E_0(R,p)$ is tight. 
A discrepancy between upper
and lower bounds on $E(R,p)$ remains for rates in the 
interval  $0<R<R_1.$ }\label{fig:rel}
\end{figure} 

In this section we study an application of the above ideas to bounds on
the function $E(R,p).$ Recently linear programming was used to
derive bounds on the distance distribution of codes \cite{lit99,ash99a}.
In particular, paper \cite{lit99} proves the following 
lower bound on the distance distribution of an arbitrary code
family of rate $R$.
\begin{theorem}{\rm \cite{lit99}} \label{thm:mu} For any family of codes of 
sufficiently large length and rate $R$ and any $\alpha\in [0,1/2]$
there exists a value $\omega, 0\le\omega\le G(\alpha,\tau)$ such that
$n^{-1}\log B_{\omega n}\ge \mu (R,\alpha,\omega)-o(1),$ where
\[
\mu (R,\alpha,\omega)= R-1+h(\tau)+2h(\alpha)-2q(\alpha,\tau,\omega/2)
-\omega-(1-\omega) h\Big(\frac{\alpha-\omega/2}{1-\omega}\Big),
\]
$\tau=h^{-1}(h(\alpha)-1+R),$ and where
\[
q(\alpha,\tau,\omega)=h(\tau)+\int_0^\omega dy
\log({P+\sqrt{P^2-4Qy^2}})/{2Q},
\]
where $P=\alpha(1-\alpha)-\tau(1-\tau)-y(1-2y), Q=(\alpha-y)(1-\alpha-y),$
is the exponent of the Hahn polynomial $H^{\alpha n}_{\tau n}(\omega
n).$
\end{theorem}
This theorem was used in \cite{lit99} to tighten the upper bound for
$E(R,p)$ for low rates, giving implicitly a condition for the union
bound to be tight for low rates. Using this result together with
Theorem \ref{thm:new}, we observe that there exists a value of the
rate $R=R_0,$ a function of $p$, such that for $0\le R\le R_0,$
the first termunder the maximum in (\ref{eq:new}) is greater than the
second one. The following statement was proved in \cite{bar04b}.
\begin{theorem}\label{thm:nb} Let $\bar R(2\rho_0(1-\rho_0))\le R_0,$
where $\rho_0$ is defined in (\ref{eq:rho0}). Then
\begin{equation}\label{eq:new-a}
E(R,p)\le -A(\bar\delta )-R+1-h(\bar\delta ) \quad 0\le R\le R_0
\end{equation}
\begin{equation}\label{eq:new-b}
E(R,p)\le\max\limits_{0\le \lambda \le \bar\delta }
\max\limits_{\lambda \le\omega\le \bar\delta }
B(\omega,\lambda)-A(\lambda)  \quad R_0 \le R.
\end{equation}
%where $A$ and $B$ are defined as in Equations {\rm(\ref{eq:Aw})} and
%{\rm (\ref{eq:Bwl})} respectively.
\end{theorem}
Explicit optimization in (\ref{eq:new-b}) is difficult
because of the cubic condition on the optimal value of the parameter $\eta$ in 
(\ref{eq:Bwl}) and for other similar reasons; however, the bound
can be computed for a given $p$. Observe that by (\ref{eq:new-a}),
for $R<R_\ast$ the BSC reliability $E(R,p)$ is estimated from above by
the exponent of the union bound. From Lemma \ref{lemma:tangent}, the
bound (\ref{eq:new-a}) is tangent on the straight-line part of
$E_0(R,p).$ 

It is clear that $R_1<\rcrit$ simply because
$\bar\delta(R)<\delta_E(R)$, i.e., the JPL function is less than the
Elias distance.
Observe that for $p\ge 0.04,$ the value $R_1\le 0.287$ (and for
$p\ge 0.05$ even $\rcrit\le 0.305$ ). 
For rates in this region we have $\bar\delta=\phi(h^{-1}(R)),$
and then the point of tangency is given by $R_1=\phi(h(\omega_{\text{typ}}))$
(since $\phi=\phi^{-1}$). 

Now to ensure that $E(R_1,p)=E_0(R,p)$ it remains to show that 
the union bound exponent can still be claimed an upper bound on $E(R,p)$
for $R=R_1,$ or that $R_1\le R_\ast.$ This can be verified by computing
the bounds (\ref{eq:new-a})-(\ref{eq:new-b}) and the value of $R_\ast.$
The computation leads to the following result (see also Fig.\,\ref{fig:rel}).

\begin{theorem} Let $p, 0.046 \le p<1/2$ be the channel transition 
probability. Then the channel reliability $E(R,p)$ equals the
random coding exponent $E_0(R,p)$ for $R_1\le R\le \rcrit.$
\end{theorem}

Previously the bound $E_0(R,p)$ was known to be tight only for
the rates $R\in[\rcrit,1-h(p)]$ \cite{eli55a}. 

Given the rate $R$ and the distance distribution of the code, the value 
of $R_\ast$ is determined uniquely. 
Based on the computational evidence, the union bound can be claimed
exponentially tight (under the approach of this section) if the code
rate satisfies (\ref{eq:condition}). Observe that Theorems 
\ref{thm:new},\ref{thm:nb} lead to the same result because of our
particular choice of the subsets $X_{ij}.$ Another possibility is
to take $\tilde X_{ij}=\{y\in X: d(x_j,y)\le d(x_i,y)\}$ in which case
these theorems would give a weaker result than \cite{coh04}
(this is the essence of the discussion in \cite[p.301]{coh04}).
%It is interesting that the results of \cite{bar04b} and \cite{coh04}
%were both first presented at the same conference (ISIT2003): while
%\cite{bar04b} guessed the correct choice, \cite{coh04} proved its
%optimality. 
The region $\tilde X_{ij}$ in Theorem \ref{thm:cm} 
is also suboptimal, but the correction term $\eta(\cdot)$
performs a transformation to the optimal region $X_{ij}$.

\section{Concluding remarks, conjectures}\label{sect:conclude}

The method of this paper and \cite{bar04b} still stops short
of proving that $E(R_0,p)$ is tight for all rates $R_x\le R\le \rcrit.$
The crucial elements of the argument made above are (a) the fact
that the JPL bound $\bar\delta(R)$ is better
than the Elias bound and (b) the straight-line
principle of \cite{sha67}.
Further progress can be related either to an
improvement of bounds on codes, which at present looks very difficult,
or to new ideas for extending a known bound on $E(R,p)$ for low rates.

We remark that the arguments and results similar to those obtained here
for the BSC can be also obtained for a power-constrained AWGN channel.
They are briefly discussed in \cite{bar04b}. The geometric picture
that describes the relation of the random coding bound and the union
bounds in this case is qualitatively the same as that of 
Sections \ref{sect:geometry}, \ref{sect:results}.

If the GV bound is tight, then so is the bound $E_0(R,p)$ on the
channel reliability. The converse claim, i.e., the implications
of the (putative) tightness of $E_0(R,p)$ for bounds on codes,
is not so obvious. To be more precise, the following question
seems open.

\medskip
\nd{\em Open problem 1.} {Assuming that the bound
(\ref{eq:Ge}b) gives an exact value of $E(R,p)$ for all $R$ in the
interval $(R_x,\rcrit),$ is it possible (with the current knowledge)
that there exists a sequence of codes whose minimum distance asymptotically
exceeds the GV distance\/}? 

\medskip\nd This is certainly not true for code sequences in which the number
of codewords of minimum weight grows subexponentially in $n$;
however, there exist codes with exponentially many minimum-weight
vectors \cite{ash01a}. A weight distribution that might
support a positive answer to the above open problem is of the form
  \begin{align*}
     B_{\omega n}&= 0 & 0<\omega<\delta\\
     B_{\omega n}&\ge 2^{n\alpha(\omega)} &  \delta<\omega,
  \end{align*}
where $\delta\ge \dgv$ and $\alpha(\omega)>R+1-h(\omega).$ Note that the
weight distribution of the code family whose existence in proved in 
\cite{ash01a} is not of this form and its distance is less that $\dgv.$
If the answer to this problem is positive, this should not be very difficult.

Given that an upper bound on $E(R,p)$ for some rate $R_0$, 
the straight-line bound of \cite{sha67} gives a method of obtaining
upper bounds on $E(R,p)$ for rates $R\ge R_0$.

\medskip
\nd{\em Open problem 2.} Given an upper bound on $E(R,p)$ for some rate
$R=R_0$ find a way of obtaining upper bounds for $R\le R_0.$

\medskip\nd This problem presently seems difficult.

So far the results for the reliability of the BSC and general discrete
memoryless channels (DMCs) 
have been similar. However, apart from straightforward
generalizations, it is not clear how to extend the result of this paper
to DMCs. Therefore, let is formulate

\medskip
\nd{\em Open problem 3.} Prove that the random coding bound on the
reliability function of a DMC is tight for rates immediately below
$\rcrit.$

\medskip\nd 
Given the similarity of results for a particular distance distribution of
Sect.\,\ref{sect:results} obtained by the methods
of \cite{cae97,coh04} and \cite{bur00,bar04b},
 another {\em open question} that arises is
whether the lower bounds of \cite{bur00} and \cite{coh04} are generally 
related.
If this is indeed the case, then the approach of \cite{coh04} would
give a more direct alternative to the successive refinement of the estimate
of $P_e(x_i)$ performed in \cite{bar04b}.
This would also have consequences in the more general context 
of hypothesis testing \cite{bur84}.

\renewcommand\baselinestretch{0.9}
{\footnotesize

\providecommand{\bysame}{\leavevmode\hbox to3em{\hrulefill}\thinspace}

}
\end{document}